\documentclass[twocolumn,pra]{revtex4}
\usepackage{amsmath,amsfonts}
\begin{document}
\title{\bf Coherent axion-photon transformations in the forward scattering on atoms}
\author{V.V. Flambaum$^{1,2,3}$, I.B. Samsonov$^{1,4}$, H.B. Tran Tan$^{1}$, D. Budker$^2$}
\affiliation{$^1$School of Physics, University of New South Wales,
Sydney 2052,  Australia,} \affiliation{$^2$Johannes
Gutenberg-Universit\"at Mainz, 55099 Mainz, Germany}
\affiliation{$^3$The New Zealand Institute for Advanced Study,
Massey University Auckland, 0632 Auckland, New Zealand}
\affiliation{$^4$Bogoliubov Laboratory of Theoretical Physics,
JINR, Dubna, Moscow region 141980, Russia}

\begin{abstract}
In certain laboratory experiments the production and/or detection of axions
is due to the photon-axion transformations
in a strong magnetic field. This process is coherent, and the rate of the
transformation is proportional to the length $l$  and magnitude $B$
of the magnetic field squared, $\sim l^2B^2$. In the present paper,
we consider coherent production of axions due to
the forward scattering of photons on atoms or molecules. This
process may be represented as being due to an effective electromagnetic field
which converts photons to axions. We present analytical
expressions for such effective magnetic and electric fields
induced by resonant atomic M0 and M1 transitions, as well as give
some numerical estimates for these fields. The corresponding
experiments would allow one to measure the electron-axion coupling
constant $g_{ae}$ in the same way as the photon-axion coupling
$g_{a\gamma}$ is studied.
\end{abstract}

\maketitle

\section{Introduction}
The axion was introduced to explain the absence of CP violation
in the strong interaction \cite{1,2,3,4,5,6,7,8,9} and is considered as a
probable candidate for a dark matter particle. The majority of experiments
searching for axions in laboratories are based on the transformation
of axions to photons (for axion detection) and photons to axions
(for axion production) in a strong magnetic field
\cite{10,11,12,13,14,15,16,17,18,19,20,21,22,23,24,25,26,27,28,29,30,31,32,33,34,35,36,37,38,39,40,42,43,44,45,46,47,48,49,50}.
The use of atomic and molecular transitions for the axion detection
has been suggested in Refs.\
\cite{Ziotas,Sikivie,Pospelov,Santamaria,Arvanitaki,Bao}.
It is natural to assume that the atomic transitions can play
role of effective electromagneic field in the process of
conversion of photons to axions.
This paper aims to estimate such
effective electric and magnetic fields induced by forward
scattering of photons on atoms near atomic resonances.

The photon-to-axion conversion probability in magnetic field $B$ is \cite{Bibber}
(in units $\hbar = c =1$)
\begin{equation}\label{Pgammaa}
P=\frac{\omega}{4k_a}\left(g_{a \gamma}B l\right)^2 F^2(q)\,,
\end{equation}
where $\omega$ is the axion and photon energy, $k_a$ is the axion momentum,
$g_{a \gamma}$ is the axion-photon coupling constant, $l$ is the
spatial extent of the magnetic field in the direction of the
axion-photon beam, $F(q) = \int{dx\, e^{-iqx} B(x)}/{\overline B} l $
is the  form factor,  $q=\omega - k_a$ is the momentum transferred
from the photon field to the axion field. For a small axion mass,
$m_a \ll \omega$, we have $q\approx 0$ and  $F(0)\approx1$.

Probabilities of the transformation of photons to axions and axions to
photons are calculated by solving the system of Maxwell and
Klein-Gordon equations coupled through the interaction term
in the Lagrangian density:
\begin{equation} \label{aEB}
L^{a \gamma} =g_{a \gamma}\, a\, ({\bf E \cdot B})\,,
\end{equation}
where $a$, $E$ and $B$ are the axion, electric and magnetic fields, correspondingly.
A similar effective Lagrangian density appears due to the interaction between the axion field and an
atom
\begin{equation}\label{EffectiveB}
L^{a \gamma}_{\rm eff} =g_{a e}\, a\, ({\bf E} \cdot {\bf B}_{\rm eff}
+{\bf E}_{\rm eff} \cdot {\bf B})\,,
\end{equation}
where the effective magnetic
${\bf B}_{\rm eff}$ and electric ${\bf E}_{\rm eff}$ fields are
produced by the atom, while $\bf E $ and $\bf B$ are physical electromagnetic fields
of an external photon. The effective Lagrangian (\ref{EffectiveB})
includes the coupling constant $g_{ae}$ which originates from the
axion-electron (or axion-nucleon) interaction
\begin{equation}\label{ae}
L^{ae}=g_{ae}\,\partial_{\mu} a\, \bar{\psi}\,\gamma^5 \gamma^{\mu} \psi
\,,
\end{equation}
where $\psi$ is the electron (or nucleon) spinor field. The aim of
this paper is to derive the expressions for the effective electric
and magnetic fields stipulated by the interaction (\ref{ae}).

The rest of the paper is organized as follows. In the next section
we derive analytical expressions for the effective magnetic and electric
fields due to the photon-axion transformation. The numerical estimates
of these fields in vapor, liquid and solid media are given in Sect.\
\ref{NumEstimates}. The obtained results are summarized in Sect.\
\ref{Concl}. In the appendices, we present the details of calculations
of the effective fields.

\section{Calculations}
Allowing for the axion-electron interaction (\ref{ae}), one can
consider a scattering amplitude, where a bound electron absorbs
an incident photon and emits an axion. Of special interest is
the scattering in forward direction since the forward scattering
on multiple atoms is always coherent.
If the wavelengths of the photon and axion are significantly larger than the distance
between the atoms, we may treat the atomic system as a continuous
medium, and the forward scattering amplitude
scales linearly with the number density of atoms $N_{\rm at}$.
Thus, the effective Lagrangian (\ref{EffectiveB})
can be identified with the amplitude of forward scattering of photons on
atoms \cite{Landau}
\begin{equation}\label{res}
L^{a \gamma}_{\rm eff}= N_{\rm at} \sum_n
\frac{M^a_{0n} M^{\gamma}_{n0}}{E_0-E_n +\omega - i \Gamma_n /2}\,,
\end{equation}
where the summation goes over the atomic (or nuclear) excited states $n$ with the energies
$E_n$ and widths $\Gamma_n$. Here $\omega$ is the energy of
the incident photon, $M^a_{0n}$ is the matrix element of the axion-electron interaction (\ref{ae}) and
$M^{\gamma}_{n0}$ is the matrix element of the interaction of
atomic electrons with external electromagnetic field of the incident
photon.

The matrix element $M^a_{0n}$ in Eq.~(\ref{res}) is given by
$M^a_{0n} = \langle 0|H^{ae}|n\rangle$, where the Hamiltonian
$H^{ae}$ corresponds to the interaction (\ref{ae}). For non-relativistic
electrons this Hamiltonian has the form \cite{Sikivie,Pospelov}
\begin{equation}
H^{ae} = g_{ae} \left(
{\boldsymbol \nabla} a \cdot \boldsymbol{\sigma}
-\partial_t a \frac{{\bf p}\cdot \boldsymbol{\sigma}}{m_e}
\right),
\end{equation}
where $m_e$ is the electron mass and $\bf p$ is its momentum. The
form of this Hamiltonian can be further specified when the axion
field is given by the plane wave with energy $\omega$ and wave-vector $\bf k$,
$a({\bf r},t) = a_0 \sin(\omega t - {\bf k}{\bf r})$. Then the Hamiltonian may be represented in the form of multipole expansion
with the leading terms given by (modulo the oscillating factor $e^{i\omega t}$)
\begin{eqnarray}
\label{H}
H^{ae} &=& g_{ae}\,a_0\,( H^{ae}_{\rm M0} + H^{ae}_{\rm M1} +\ldots)\,,\\
H^{ae}_{\rm M0}  &=& \frac12 \left(-\frac{\omega}{m_e} {\bf p}\cdot\boldsymbol{\sigma}
+i ({\bf k \cdot r} )({\bf k } \cdot \boldsymbol{\sigma})\right)\,,\\
H^{ae}_{\rm M1} &=& -\frac12 ({\bf k}\cdot \boldsymbol{\sigma})\,.
\end{eqnarray}

The scattering amplitude (\ref{res}) may be considered as a
process, where an atom is (virtually) excited by absorbing a photon
and returns back to the initial state via axion emission. Note
that the M0 axions induce transitions between the states of
opposite atomic parity but the same total angular
momentum. Thus, M0 axions are produced when atoms absorb E1
photon which also changes the parity. Analogously, M1
axions do not change parity, and they may be produced when M1 photons
are absorbed.
As a result, in Eq.\ (\ref{res}) the matrix element of the
operator $H^{ae}_{\rm M0}$ should be coupled with the
matrix element of E1 photon absorption operator $H^\gamma_{\rm
E1}=e (\boldsymbol{\epsilon}\cdot {\bf r})$, while the matrix element of the operator $H^{ae}_{\rm M1}$
should be multiplied by the matrix element of M1 photon-absorption
operator $H^\gamma_{\rm M1}=\frac{e}{2m_e}({\bf n}\times \boldsymbol{\epsilon})
({\bf J}+{\bf S})$, where ${\bf n}$ is the propagation unit vector of the photon and
$\boldsymbol{\epsilon}$ is its polarization. The former case corresponds to
the effective magnetic field while the latter generates the effective
electric field due to the interaction of photons with atoms,
\begin{eqnarray}
B_{\rm eff} &=& N_{\rm at}
{\rm Re} \sum_n \frac{\langle 0| H^{ae}_{\rm M0}| n \rangle
\langle n |H^{\gamma}_{\rm E1}|0\rangle }{E_0 - E_n + \omega -i \Gamma_n
/2}\,,\label{Beff}\\
E_{\rm eff} &=& N_{\rm at}
{\rm Re} \sum_n \frac{\langle 0| H^{ae}_{\rm M1}| n \rangle
\langle n |H^\gamma_{\rm M1}|0\rangle }{E_0 - E_n + \omega - i \Gamma_n
/2}\,.\label{Eeff}
\end{eqnarray}

The direction of the effective fields is specified by the matrix
elements in Eqs.\ (\ref{Beff}) and (\ref{Eeff}). In fact, as we
prove in our previous paper \cite{Bao}, for non-polarized atoms
the product of axion and photon matrix elements always sums to
zero, $\sum_n\langle 0| H^{ae}| n \rangle
\langle n |H^{\gamma}|0\rangle =0$, because the
excited states $|n\rangle$ with different projections of the total
momentum contribute destructively. To get a non-vanishing result
we have to assume that a (weak) external magnetic field $B_2$ is
applied, which lifts the degeneracy of Zeeman sublevels such that
only some of them are in resonance with the applied laser field.
In this case, it is possible to show that the direction of the effective
field $B_{\rm eff}$ coincides with the direction of the applied
physical magnetic field $B_2$ while ${\bf E}_{\rm eff}$ is directed
along ${\bf k} \times {\bf B}_2$.

A resonant atomic transition may lead to a significant photon absorption and a phase shift of the photon field (relative to the axions field) which destroys the coherence.
Both effects may be seen from the complex phase in the photon field
$A  \propto e^{i n_r {\bf k \cdot r }}$, where
\begin{equation}
n_r = 1-2\pi N_{\rm at} \sum_n \frac{|M^{\gamma}_{n0}|^2}{E_0-E_n +\omega  - i \Gamma_n /2}
\end{equation}
is the complex refractive index. One could go to the tail of
resonance to suppress the photon absorption since the imaginary
part of $n_r$ decreases quadratically with $E_0-E_n +\omega$,
but the phase decreases slowly (linearly), similar to $ E_{\rm eff}$ and  $ B_{\rm eff}$
in Eqs.\ (\ref{Beff}) and (\ref{Eeff}). The phase of the photons may be
corrected by placing phase-correcting transparent plates in the
medium. This situation is opposite to the case of photon-axion
transformation in magnetic field \cite{vanBibber}, where it was
proposed to use a gas to match the photons' speed with the speed
of low-mass axion. In our case the phase-correcting transparent
plates may be needed if the photons' phase velocity in the medium appears
less than the axion's speed. In this case the plate should add
the photons' phase to equate the phase difference with axions to a
multiple of $2 \pi$.

\section{Numerical estimates}
\label{NumEstimates}
Note that the effective fields (\ref{Beff}) and (\ref{Eeff}) are
proportional to the atom density $N_{\rm at}$. Thus, a stronger
effective field is produced in dense atomic systems such as
solids, liquids or dense gases. In this note we consider
noble gases in liquid phase and heavy metal vapors since they can be treated using
the standards atomic methods. We also show that there is a
significant enhancements of the effective field in crystals with
narrow spectral lines.

\subsection{Effective magnetic field in liquid xenon}
Let us consider liquid xenon with atomic density $N_{\rm at}=1.3\times
10^{22}$ cm$^{-3}$. The resonant axion M0 transition is possible from the ground
state $0^+$ to the excited state $0^-$ with the energy
$E_{0^-}=9.45$ eV, while the corresponding photon transition is
highly forbidden. This photon transition may be open when
an external magnetic field $B_2$ is applied. This magnetic
field causes the mixing of the states $0^-$ and $1^-$, where the
latter corresponds to the energy level $E_{1^-}=9.57$ eV. As we
demonstrate in Appendix \ref{AppA}, the effective magnetic field
may reach the value
\begin{equation}
B_{\rm eff}\approx 1.2\times 10^{-2}
\ {\rm T}.
\label{Beff1}
\end{equation}
We stress that, according to Eq.\ (\ref{EffectiveB}), the strength of interaction of this effective field
with the axion field is specified by the coupling constant
$g_{ae}$, which is independent from the axion-photon constant
$g_{a\gamma}$. Thus, although this effective field is much weaker than the physical
magnetic field $B=5.3$ T applied at the ALPS experiment
\cite{ALPS}, it may be significant in the processes of axion
production or detection if $g_{ae}\gg g_{a\gamma}$.

\subsection{Effective electric field in vapor thallium}
To estimate the effective electric field (\ref{Eeff}) we consider
a vapor of Tl atoms at temperature $T=1200$ $^{\circ}$C and
density $N_{\rm at}=6.6\times 10^{17}$ cm$^{-3}$. This system
proved useful in the study of weak-interaction-induced optical activity of heavy-metal
vapors \cite{Khriplovich}. Indeed, the transition between the
ground state $A$ with $J=\frac12$ and the nearest excited state
$B$ with $J=\frac32$ is of M1 type. The energy difference between
these states $\omega=0.966$ eV corresponds to near-infrared region.
In Appendix \ref{AppB} we give an estimate of the effective
electric field corresponding to such atomic transitions:
\begin{equation}
E_{\rm eff}\approx 1.5\times 10^{5} \, {\rm \frac{V}{cm}}\,.
\label{Eeff1}
\end{equation}
This field is still not competitive with the physical magnetic field
used at ALPS experiments.

\subsection{Effective electric field in a crystal}
\label{crystal}
According to equations (\ref{Beff}) and (\ref{Eeff}), the
strong effective fields are achieved in media with high atom
density $N_{\rm at}$ and small width $\Gamma$. Therefore it is
natural to consider photon-to-axion transformations in crystals
with narrow optical lines. One of such crystals based on
EuCl$_3\cdot 6$H$_2$O compound has recently been studied in \cite{Sellars}.
Remarkable spectral properties of this crystal are due to the 4f valence
electron shell of the Eu$^{3+}$ ion \cite{Binnemans,Ofelt}.
In particular, in \cite{Sellars} it was shown that the optical
$^7F_0\to {}^5D_0$ transition in such isotopically purified crystal may be as narrow as 25
MHz. However, we cannot employ this transition for axion
production since it corresponds to the induced electric dipole type
and does not obey the axion-amplitude selection rules. Instead, we
consider the transition $^7F_0\to {}^5D_1$ with the energy $\omega=2.36$ eV in the Eu$^{3+}$
ion, which is of M1 type \cite{Binnemans} and is allowed for both the photon and axion
amplitudes. Note that the states $^7F_0$ and $^5D_1$ refer to the intermediate
coupling scheme, and each of them is given by a linear combination
of the LS-states according to \cite{Ofelt}.

Although there is no explicit data about the width of the $^7F_0\to {}^5D_1$
transition in the EuCl$_3\cdot 6$H$_2$O crystal, we assume that it
may be as narrow as for the $^7F_0\to {}^5D_0$ transition, $\Gamma \sim 10^{-7}$ eV.
Then, taking into account that the density of
Eu atoms in this crystal is $N_{\rm at}\approx 2.7\times
10^{21}$ cm$^{-3}$ \cite{Sellars}, in Appendix \ref{AppC} we
estimate the effective electric field due to the photon-axion
transformation:
\begin{equation}
E_{\rm eff} \approx 1.7 \times 10^{9}  \ \frac{\rm V}{\rm cm}\,.
\label{Eeff2}
\end{equation}
Thus, the effective field in crystals (\ref{Eeff2}) is
much stronger than that in vapor media (\ref{Eeff1}) due to higher atom density
and smaller linewidth.

We point out that Eq.\ (\ref{Eeff2}) gives a rough
estimate of the effective field because we do not know the exact
width of the line corresponding to the $^7F_0\to
{}^5D_1$ transition in the EuCl$_3\cdot 6$H$_2$O crystal. We hope that a more
accurate estimate may be given when this transition is
experimentally measured.

\section{Concluding remarks}
\label{Concl}
The aim of this note is to attract the attention to the
possibility to have resonance enhancement of the axion-photon
transformations in a medium. Such transitions between photons and
axions in atoms can be represented via the effective electric
and/or magnetic fields, similar to the interaction of the axion
with physical electromagnetic fields. The crucial difference of
these effective fields from the physical ones is that they couple
to the axion through the independent coupling constant $g_{ae}$
which, for general axion-like particles, may be much larger than the axion-photon
coupling $g_{a\gamma}$. This motivates to set up new axion
production/detection experiments where the medium plays the role
of physical magnetic field in the current axion-search
experiments. Such new experiments would allow one to measure the
axion-electron coupling constant in the same way as the
axion-photon coupling is studied. It is tempting to design such
experiments and to estimate their efficiency.

We point out that the equations (\ref{Beff}) and (\ref{Eeff}) give
an important hint towards the design of future atom- and
molecule-based detectors with high sensitivity to axion-like
particles: They should possess narrow spectral lines at relatively
high atom density. In particular, it is natural to look for
crystals with ultranarrow optical or microwave transitions.
Examples of such crystals with the linewidth $\Gamma\sim 10^{-7}$
eV in the optical region have recently been studied, see, e.g.,
\cite{Sellars}. In Sect.\ \ref{crystal}
we demonstrated that the use of such
crystals as a media for photon-axion transformation may give a
significant enhancement of the effective field which may be much
stronger than the physical fields employed at the ALPS
experiments, see, e.g., \cite{Breview} for a review. It is
tempting to design such detectors and estimate their efficiency
more accurately for applications in future ALPS experiments. We
leave this issue for further work.

\vspace{3mm}
{\bf Acknowledgments.}
    The authors thank R. Ahlefeldt, V. Dzuba,  C. Rizzo, M. Sellars, and Y. Stadnik for useful
    discussions. This work is supported by the Australian Research Council,
    the Gutenberg Fellowship, New Zealand Institute for Advanced
    Studies, the DFG via its Reinhardt Koselleck Program,
    and the European Research Council under the European Union's
    Horizon 2020 Research 41 and Innovative Program under Grant agreement No.~695405.

\appendix
\section{Estimates of effective fields}

\subsection{Effective magnetic field produced by
liquid xenon}\label{AppA}

Let us consider the excited state $0^-$ in Xe atom with
energy $E_{0^-}\equiv\omega=9.447$ eV. This state has zero total angular momentum,
$J=0$, and odd parity in contrast with the even-parity ground state $0^+$. Therefore, the axion transition between these
states is of M0 type. The corresponding matrix element for resonant axion absorption
was calculated in our recent work \cite{Bao}
\begin{equation}
M^a =\langle 0^- | H^{ae}_{\rm M0} | 0^+ \rangle
= i\frac{\sqrt2}3  \omega^{2} R\,,
\label{10}
\end{equation}
where $R$ is the radial integral
$R=\int_0^\infty
f_{6s_{1/2}}(r)f_{5p_{1/2}}(r)r^3 dr$.
Here $f_{6s_{1/2}}(r)$ and $f_{5p_{1/2}}(r)$ are radial parts of $6s_{1/2}$ and $5p_{1/2}$
wavefunctions, respectively (see, e.g.\ \cite{Khriplovich}).
For Xe atom, the value of this
integral may be deduced from \cite{NIST}, $R\approx - 1.09$ a.u.

The photon transition $0^+\leftrightarrow 0^-$ is highly
forbidden. However, this transition may be opened by applying a
magnetic field $B_2$ which provides the mixing of the excited states $|0^-\rangle$ and
$|1^-\rangle$,
\begin{equation}
|i\rangle = |0^-\rangle
-\frac{\langle 1^-|\boldsymbol{\mu} \cdot {\bf B}_2 |0^-\rangle}{E_{0^-} -
E_{1^-}}|1^-\rangle\,,
\label{11}
\end{equation}
where $E_{1^-}=9.57$ eV; $\boldsymbol{\mu}=-\mu_B({\bf J}+{\bf S})$ is the electron magnetic moment operator
and $\mu_B$ is the Bohr magneton.
Taking into account that the matrix element $\langle 0^-|\boldsymbol{\mu} \cdot {\bf B}_2
|1^-\rangle$ is $2/3\mu_B B_2$, the photon E1 transition amplitude
between the ground state $|0^+\rangle$ and the excited state $|i\rangle$ is
found to be
\begin{equation}
M^\gamma = \frac 23 \frac{ e \mu_B B_2 }{E_{0^-} -
E_{1^-}}\langle 1^- | \boldsymbol{\epsilon}\cdot {\bf r} |
0^+ \rangle
=\frac{\sqrt2}{9}\frac{e \mu_B B_2 R}{E_{0^-} -
E_{1^-}}
\,,
\label{12}
\end{equation}
where $\boldsymbol{\epsilon}$ is the photon polarization vector
which is assumed to be along the $z$-axis, $\boldsymbol{\epsilon}=(0,0,1)$.
Thus, for the product of transition amplitudes
(\ref{10}) and (\ref{12}) we find
\begin{equation}
M^a M^\gamma = \frac{2i}{27} \frac{e \mu_B B_2}{E_{0^-} -
E_{1^-}}\omega^2 R^2\,.
\label{13}
\end{equation}

Note that the expression (\ref{13}) is imaginary. Hence, the real
part of the effective Lagrangian (\ref{res}) corresponds
to the resonant absorption. Note also that
for liquid Xe the dominant
contribution to the width of the state $\Gamma$ comes from the
collisional broadening, $\Gamma\approx \Gamma_{\rm col} = 2 v_0 N_{\rm at}
\sigma_{\rm col}$, where $v_0=\sqrt{2 k_B T/m_{\rm at}}$ is the most
probable thermal speed of the atoms and $\sigma_{\rm col}$
is the collisional cross section. Here $k_B$ is the
Boltzmann constant and $m_{\rm at}$ is the atomic mass.
Assuming that the temperature of liquid
Xe is $T=164$ K and the atom density is $N_{\rm at} = 1.3\times
10^{22}$ cm$^{-3}$, the collisional width is estimated as
$\Gamma_{\rm col}\approx 1.4 \times 10^{-3}$ eV.
Taking into account Eq.\ (\ref{13}), we find the expression for the
effective magnetic field
\begin{equation}\label{A5}
B_{\rm eff} = \frac2{27} N_{\rm at}
\frac{\mu_B B_2}{E_{0^-} -
E_{1^-}} \frac{\omega^2 R^2 }{\Gamma_{\rm col}}\,.
\end{equation}

Assuming that the magnetic field $B_2$ is of order of the magnetic
field applied at ALPS experiment \cite{ALPS}, $B_2=5$ T,
we estimate the effective magnetic field produced by the atomic
transitions (\ref{A5}):
$B_{\rm eff} \approx 1.2\times 10^{-2}$ T.

\subsection{Effective electric field produced by
thallium vapor}\label{AppB}

Consider the ground and excited states of the Tl atom, $A =
|L=1,J=\frac12\rangle$ and $B = |L=1, J=\frac32\rangle$,
with energy difference $\omega= 0.966$ eV. The transition between
these states is of M1 type. The axion and photon absorption
amplitudes for this transition are
\begin{eqnarray}
M^{a} &=& -\frac12
\langle B | {\bf k}\cdot \boldsymbol{\sigma} |A\rangle\,,\label{21}\\
M^\gamma &=& \frac{e}{2m_e}  ({\bf n}\times \boldsymbol{\epsilon})
\langle B | {\bf J}+{\bf S} |A \rangle\,,\label{22}
\end{eqnarray}
where $\bf n$ is the photon or axion propagation unit vector and
$\boldsymbol{\epsilon}$ is the photon polarization vector.
For unpolarized atoms, in (\ref{Eeff}) we have to average over
the states with different $z$-projection of the total angular
momentum $M_J$. After
this averaging, the contribution to the effective electric field
vanishes, $\sum_{M_J} M^a M^\gamma =0$, see \cite{Bao}.

To get a non-vanishing result in (\ref{Eeff}) one can apply a
magnetic field $B_2$, which splits the Zeeman sublevels. The laser
frequency is in resonance with a transition from only one of these
sublevels. In particular, we suppose that the constant magnetic
field $B_2$ is applied along the $z$-axis, ${\bf B}_2 = B_2 \hat {\bf
z}$, and the laser frequency corresponds to the energy difference of
levels $A= |1,\frac12,-\frac12 \rangle$ and
$B=|1,\frac32,\frac12\rangle$. We assume also that the photon
propagates along the $y$-axis and is polarized in the
$(x,z)$-plain, ${\bf n}=(0,1,0)$,
$\boldsymbol{\epsilon}=(\epsilon_x,0,\epsilon_z)$. In this case,
the axion and photon matrix elements (\ref{21}) and (\ref{22}) are
\begin{eqnarray}
M^a &=&\frac i{3\sqrt2} \omega I\,,\\
M^\gamma&=& \frac{e \epsilon_z I}{6\sqrt2m_e}\,,
\end{eqnarray}
where $I$ is the overlap integral for $6p_{1/2}$ and $6p_{3/2}$
wavefunctions, $I=\int_0^\infty dr\, r^2 f_{6p_{1/2}}(r)f_{6p_{3/2}}(r)\approx
0.98$.

As is shown in \cite{Khriplovich}, for Tl vapor at a temperature of $T=1473$ K, the dominant
contribution to the width of the line comes from the Doppler
broadening, $\Gamma \approx \Gamma_{\rm Dop}=2 v_0 \omega/\sqrt\pi
$, where $v_0=\sqrt{2 k_B T/m_{\rm at}}$ is the most probable thermal
speed of the atoms. The numerical estimate for this width is
$\Gamma_{\rm Dop}\approx 1.3\times 10^{-6}$ eV. Taking this into
account, we have the following expression for the effective
electric field
\begin{equation}
E_{\rm eff} = N_{\rm at }\frac{e }{18 m_e }
\frac{\omega I^2\epsilon_z}{\Gamma_{\rm Dop}}\,,
\label{25}
\end{equation}
where the density of the vapor is $N_{\rm at}=6.6\times 10^{17}$
cm$^{-3}$.

Assuming that the polarization of the photon is such that
$\epsilon_z=1$, we estimate the effective electric field
$ E_{\rm eff}\approx 1.5\times 10^{5} $
${\rm \frac{V}{cm}}$.

Note that the same result may be obtained using the optical
pumping rather than applying the magnetic field $B_2$.

\subsection{Effective electric field in crystals}
\label{AppC}
Let us consider a EuCl$_3\cdot$6H$_2$O crystal, where there are
narrow spectral lines in the optical region, and high optical density can be achieved, see
\cite{Sellars}. The optical properties of this crystal can
be understood through the electron-shell structure of the
Eu$^{3+}$ ion which has six valence electrons on the 4f orbital \cite{Binnemans,Ofelt}.
We are interested in the transition between the states
$A={}^7F_0$ and $B={}^5D_1$ corresponding to the energy
$\omega=2.36$ eV. Since this transition is of M1 type, we apply
the formula (\ref{Eeff}) to estimate the effective electric field $E_{\rm eff}$ due
to axion-photon transformation.

The relevant computation is similar to the one given in Sect.\
\ref{AppB} for Tl vapor. In particular, it is also necessary to apply the
constant magnetic field $B_2$ along the $z$-axis to lift the
degeneracy of the Zeeman sub-levels of state $B={}^5D_1$ with the
total angular momentum $J=1$. To be specific, we assume
that the laser frequency is in resonance with the transition
between the states $|A\rangle = | {}^7F_0; J=0,M=0\rangle$ and
$|B\rangle = | {}^5D_1; J=1,M=1\rangle$. For simplicity, we also assume
that the laser beam is aligned along the $y$-direction and is polarized in the
$z$-direction, so that ${\bf n}=(0,1,0)$ and
$\boldsymbol{\epsilon}=(0,0,1)$. In this case, using the
equations (\ref{21}) and (\ref{22}) we find
\begin{eqnarray}
M^a &=& -\frac i{\sqrt6} \omega \langle {}^5D_1 || S || {}^7F_0
\rangle\,,\\
M^\gamma &=& -\frac{e}{2\sqrt6 m_e}
\langle {}^5D_1 || J + S || {}^7F_0 \rangle\,,
\end{eqnarray}
where $\langle {}^5D_1 || S || {}^7F_0 \rangle$ is the reduced
matrix element of the spin operator $\bf S$ and $\langle {}^5D_1 || J + S || {}^7F_0 \rangle
=\langle {}^5D_1 || S || {}^7F_0 \rangle$. This matrix element
may be computed using the spectral data for the Eu$^{3+}$ ion
\cite{Binnemans,Ofelt}: $\langle {}^5D_1 || S || {}^7F_0
\rangle\approx0.17$.

To apply the equation (\ref{Eeff}) we need the
atom density $N_{\rm at}$ and the width of the state $\Gamma$. The
former may be easily estimated for the crystal
EuCl$_3\cdot$6H$_2$O described in \cite{Sellars}: $N_{\rm at}\approx 2.7\times
10^{21}$ cm$^{-3}$. The authors of this paper claim that the width
of the optical transition $^7F_0 \to {}^5D_0$ in this crystal is
about 25 MHz. For our estimates, we can assume that the transition
$^7F_0 \to {}^5D_1$ possesses a comparable width $\Gamma \sim
10^{-7}$ eV. For these values of the parameters, we find the
effective electric field in the crystal due to the photon-axion
transformation,
\begin{equation}
E_{\rm eff}=\frac{N_{\rm at}e \omega }{12 m_e \Gamma}|\langle {}^5D_1 || S || {}^7F_0
\rangle|^2 \approx 1.7\times 10^{9} \frac{\rm V}{\rm cm}\,.
\end{equation}
We point out that these estimates are very rough; a more accurate
estimate requires experimental measurements of the width of the
line for the transition $^7F_0 \to {}^5D_1$ in the
EuCl$_3\cdot$6H$_2$O crystal.

\end{document}